\documentclass[preprint, aps,amsmath,amssymb,floats,12pt]{revtex4}
\usepackage{amsmath}
\usepackage{graphicx}
\usepackage{dcolumn}
\usepackage{bm}
\usepackage[dvips]{color}
\begin{document}

\title{MAXENT and the Tsallis Parameter}
\author{ J. M. Conroy\thanks{Justin.Conroy@fredonia.edu} and H. G.
Miller\thanks{E-mail:hgmiller@localnet.com}}
\affiliation{Physics Department, The State University of New York at Fredonia, 
Fredonia, NY, USA}

\begin{abstract}
The nonextensive entropic measure proposed by Tsallis introduces a parameter, 
q, which is not defined but rather must be determined.  The 
value of q is typically determined from a piece of data and 
then fixed over the range of interest.  On the other hand, from a 
phenomenological 
viewpoint,  there are  instances in which q cannot be treated as a constant. 
 We present two distinct approaches for determining q depending on the form of 
the equations of constraint for the particular system.  In the first case the 
equations of constraint for an operator O can be written as $Tr[F^{q}O]=C$, 
where C may be an explicit function of the distribution function, F.   In this 
case 
one can solve an equivalent MAXENT problem which yields q as a function of the 
corresponding 
Lagrange Multiplier.  As an illustration the exact solutions to the static 
Generalized Fokker-Planck Equation (GFP) are obtained from MAXENT.   
As in the case where C is a constant if
 q is treated as a variable within the MAXENT 
framework, the entropic measure is maximized for all values of q trivially.  
Therefore q must be determined from existing data. In the 
second case an additional 
equation of 
constraint exists which cannot be brought into the above form.  In this case 
the additional equation of constraint may be used to determine the fixed value 
of q. 
\end{abstract}
\maketitle

Since its introduction the Tsallis entropy\cite{T88,IT89,AO01,PP99,PMP04} has  
increasingly 
been 
utilized as the entropic measure in the Maximum Entropy (MAXENT) 
caculations\cite{J03}. 
The choice of the Tsallis parameter, q, which is not defined {\it a 
priori}\cite{T99,T09a} can yield in the limit {q $\rightarrow$ 1} the 
Botzmann-Gibbs (BG) entropy as well as a family of fractal entropies (q$\neq$ 
1). It
has been argued that perhaps this parameter should be constant, if not
universally, at least for classes of dynamical systems\cite{GT04,WW00,TBT07} 
and 
attempts have been made to set limits on the value of this parameter \cite
{TBIB}. 
Within the framework of the MAXENT approach the question arises as to how one 
should determine q. 

For a classical system the stationary state distribution function, F, can 
 be obtained from the classical
Tsallis entropy 
\begin{equation}
S_{q}=\frac{1}{q-1}(F-F^{q})
\end{equation}
via the  MAXENT equation\cite{PP99}
\begin{equation}
 \delta_F S_q =0
\label{Sq}
\end{equation}
along with the equation of constraint
\begin{equation}
Tr[F^{q}O]=C
\label{eoc}
\end{equation}
The solution of these equations yields the classical Tsallis distribution given 
by

\begin{equation}
F^q(x)=D[1-\beta(1-q)O(x)]^{\frac{1}{1-q}},
\label{Fq}
\end{equation}

where D is a constant.
We consider two cases. 

In the first case all of the equations of constraint are 
of the form of equation(\ref{eoc}).  We consider the following non-linear 
one dimensional Generalized Fokker-Planck (GFP) equation\cite{PP95}
\begin{equation}
\frac{\partial F}{\partial t}=-\frac{\partial}{\partial
x}\{K(x)F\}+\frac{1}{2}Q\frac{\partial^{2}[F^{2-q}]}{\partial x^{2}},
\end{equation}
where F is the distribution function, Q is the diffusion coefficient, and K(x)
is the drift coefficient which determines the potential:
\begin{equation}
V(x)=-\int_{x_{0}}^{x} K(x)dx.
\end{equation}
The particular power $q-2$ is chosen in accordance with the quite general
discussion of the
generalized Bogulubov inequalities\cite{PT93} which points out that systems
which obey Tsallis statistics exhibit abrupt changes at $q=2$.
 An exact solution (both static and time dependent) of the GFP equations 
has been found and is shown under
certain circumstances to be equivalent to the Tsallis classical distribution
functions\cite{PP99}.

 In the static case we formulate and solve the equivalent MAXENT
problem where 
the $C$ is an explicit 
function of the distribution functions.  In the GFP case one has
\begin{equation}
 K(x)F=1/2 Q\frac{\partial F^{2-q}}{\partial x}
\label{pp}
\end{equation}
\begin{equation}
\int^x KFdx= Q/2 (F^{2-q} (x) - F^{2-q} (x_0))
\end{equation}
Integrating by parts (K=-$\frac{\partial V}{\partial x}$ and V($x_0$) = 0) 
\begin{equation}
V(x)F(x)=-Q/2 (F^{2-q}(x) -F^{2-q}(x_0) +\int V\frac{\partial F}{\partial x})
\end{equation}
or

\begin{equation}
tr[VF^q]=tr[F^{q-1}(\int V\frac{\partial F}{\partial 
x}dx-Q/2(F^{2-q}(x)-F^{2-q}(x_0))]
\label{ec}
\end{equation}
Now eq(\ref{pp}) is solved by the solution given in the 
Plastinos paper
\begin{equation}
F(x)=D[1-\beta(1-q)V(x)]^{\frac{1}{1-q}}.
\label{Fq1}
\end{equation}
One can
verify by substitution  that Eq.(\ref{Fq})  is a solution to Eq. (\ref{pp})
 provided
\begin{equation}
\beta=\frac{2}{Q}[\frac{D^{q-1}}{2-q}]
\label{beta}
\end{equation}
With D=1 the solution is the same as that 
obtained from the above MAXENT equations (with eq(\ref{ec}) as the equation of 
constraint). Again the equation of constraint is only satisfied  if $\beta$ is 
given by  equation(\ref{beta}) (with D$=$1).
 Note in this case there is no solution if V(x) 
is a 
constant.

Generally the C in the equation of constraint is a 
constant rather than an explicit function of the distribution finctions.
In order to simultaneously
determine q it has been suggested \cite{PMP04} that an additional equation 
\begin{equation}
 \frac{\partial S}{\partial q}|_{\beta}=0
\label{dsdq}
\end{equation}
must be solved. However it should be noted that this equation can be re-written 
as
\begin{eqnarray*}
 \frac{\partial S}{\partial q}|_{\beta}&=& \frac{\partial S}{\partial 
F}|_{\beta} \frac{\partial F}{\partial q}|_{\beta}\\
            &=& 0.
\end{eqnarray*}
Since Tsallis distribution functions (equation(\ref{Fq})) are the solution of 
equations(\ref{Sq}) and (\ref{eoc}) the above equation is trivally satisfied 
for any value of q. Hence q cannot be determined in this manner and one has no 
choice but to determine q from the existing data. For the practitioners this 
has been accepted de facto and calculations 
involving the
Tsallis entropy have generally used one piece of data  to determine the value of
q which
is then fixed over the range of interest\cite{TBIB}.  

In the second  case an additional equation of constraint exists which cannot be 
brought into the form of equation (\ref{eoc}). Such is the case in obtaining 
the distribution functions of the  finite temperature BCS equations with the 
Tsallis single particle entropic measure.  Aside from the self consistency 
requirements of the single quasi-particle energies, which are needed, the 
determination of the distribution functions follows from equation(\ref{Sq}) 
along with the appropriate equations of constraint. The existence of a critical 
point where the gap vanishes yields an additonal   equation of 
constraint not of the form given by equation (\ref{eoc}). As has been shown 
recently this,  can be used to determine 
q\cite{UMK01,CM08}.



\end{document}